\documentstyle[prc,aps,axodraw,preprint]{revtex}
\begin{document}

\title{ {\bf \Large Gauge invariance in quantum hadrodynamics } }

\author{ {\large Gary Pr\'{e}zeau}\\ {\it Department of Physics, College of 
William{\&}Mary, Williamsburg, Virginia 23187-8795} }


\maketitle

\begin{abstract}

This paper describes the derivation of a  new
representation of an $SU(2)_L \times SU(2)_R$ locally-invariant model
of the strong interactions called quantum hadrodynamics 3 (QHD-III).
QHD-III is a gauge-invariant theory based on the linear $\sigma$ model,
the gauge bosons being the $\mbox{\boldmath $\rho$}$ and the $\mbox{\boldmath $a_1$}$.  The new
representation considerably simplifies the lagrangian.  To derive the
new representation, the gauge invariance of the model is exploited.  The
role of the gauge boson masses in 
gauge-invariant models of the strong interactions based on the
$\sigma$ model is also discussed, and it is shown that these masses are necessary
if the pion is to survive as a physical field.

\end{abstract}
\vspace{20pt}

Hadronic models have had significant phenomenological success in
describing the many-body  strongly
interacting system at low energies \cite{walecka}.  These models take hadrons as their 
effective degrees of freedom.  The lagrangians of hadronic models are
constructed in such a way that they reflect
the symmetries of QCD, while incorporating low-energy phenomenology.
In particular, these models must give rise to conserved vector and
partially conserved axial-vector currents, and they should include the
exchange of mesons which are known to carry the strong
force at low energies.  The mesons should be introduced such
that corrections coming from meson loops are consistently calculable within the model.
Since there is a large number of mesons, any model
based on meson exchange must find a consistent way to choose the
relevant mesons; this can be achieved by introducing them as gauge
bosons.

Briefly, QHD-III \cite{qhd3} is a gauge-invariant
 hadronic quantum field theory based
on the gauged $\sigma-\omega$ model with pions.  The $\sigma-\omega$ model is built
from the linear $\sigma$ model with the $\omega$ introduced as a
massive $U(1)$ gauge boson (see below).  When the global $SU(2)_L\times SU(2)_R$
symmetry of the $\sigma-\omega$ model is made local and parity
conservation is imposed, the 
$\mbox{\boldmath $\rho$}$ and its chiral partner, the $\mbox{\boldmath $a_1$}$, appear as the gauge bosons.
The $\mbox{\boldmath $\rho$}$ and the $\mbox{\boldmath $a_1$}$ are made massive by the inclusion of a Higgs
sector composed of two complex doublets: one transforming under
$SU(2)_L$
and the other under $SU(2)_R$;
the doublets couple to the gauge bosons through their
covariant derivatives.  This procedure for giving mass to the $\mbox{\boldmath $\rho$}$
and the $\mbox{\boldmath $a_1$}$ is very similar
to the one used to give mass to the weak bosons in the standard model,
and preserves the gauge invariance of the model.  Keeping the gauge
invariance allows the unambiguous derivation of the strong conserved currents.
A small symmetry-breaking term is included to yield massive pions.  The
resulting lagrangian has a minimal number of massive mesons which
couple to conserved vector and axial-vector currents; in this model, the $\mbox{\boldmath $a_1$}$
naturally comes out heavier than
the $\mbox{\boldmath $\rho$}$.  The lagrangian is also renormalizable \cite{qhd3}.

As shown below, the physical pion disappears in an hadronic model based on the
gauged $\sigma$ model.\footnote{This point has also been previously independently noted in
\cite{meissner}}  To retain the pion, the gauge bosons must be given mass.
In contrast to QHD-III, the usual procedure
\cite{julian,ga,zuddin,ko} is to
put in the same mass, $m_\rho$, by hand for both the $\mbox{\boldmath $\rho$}$ and the $\mbox{\boldmath $a_1$}$.
The spontaneous symmetry-breaking (SSB) that occurs in the $\sigma$ sector provides an
extra contribution to the mass of the $\mbox{\boldmath $a_1$}$ making the $\mbox{\boldmath $a_1$}$ mass, $m_a$,
larger than the $\mbox{\boldmath $\rho$}$ mass, $m_\rho$.  However, introducing mass terms for the
gauge bosons by hand violates current conservation, and loops can no
longer be calculated unambiguously.  Furthermore, because of the SSB in the $\sigma$
sector, the $\mbox{\boldmath $a_1$}$ and the gradient of the pion mix, and the resulting
lagrangian must be diagonalized.  The diagonalization of the
lagrangian is carried out by making a change of variables involving
the $\mbox{\boldmath $a_1$}$ and the gradient of the pion.  The final lagrangian is
complicated because of the introduction of momentum-dependent vertices 
due to the gradient of the pion.

In a gauge-invariant quantum field theory such as QHD-III \cite{qhd3}, 
one can make a gauge transformation to diagonalize the lagrangian
instead of making the above-mentioned change of variables.  The 
result is a considerably simpler lagrangian where no new momentum
dependent vertices appear.  In particular, to ${\cal{O}}(g_\rho^2)$,
the diagonalization of the lagrangian is equivalent to the rescaling of the
pion field by the ratio $m_\rho / m_a$.  This work details the
derivation of this new representation of the QHD-III lagrangian.

This paper begins by discussing the $SU(2)_L\times SU(2)_R$ locally
invariant $\sigma-\omega$ model with pions.  The gauge bosons, the
$\mbox{\boldmath $a_1$}$ and the $\mbox{\boldmath $\rho$}$, are originally massless.  It is shown that the
SSB gives a mass to the $\mbox{\boldmath $a_1$}$.  Working in an
arbitrary $\xi$ gauge, it is shown that the field originally
identified with the pion acquires a $\xi$-dependent mass, which
identifies it as a {\it fictitious particle}; what looks like a pion is
labeled as $\mbox{\boldmath $\pi$}^\prime$ to distinguish it from the physical pion.  By
looking at nucleon-nucleon scattering, it is shown that the
{\boldmath $\pi^\prime$} exchange diagram is always {\it canceled} by the $\xi$
dependent part of the $\mbox{\boldmath $a_1$}$ exchange diagram.
The disappearance of the real pion is forced by gauge invariance and
demonstrates the need for massive gauge
bosons with mass provided from outside this sector of the theory.  

QHD-III is then reviewed.  In this locally gauge-invariant model,
pions appear as the physical Goldstone bosons.  Here, the vector meson masses are generated
from a Higgs sector, as in the $\sigma$ model.  Gauge invariance is used to diagonalize the
lagrangian in an arbitrary $\xi$ gauge so as to avoid the original,
momentum-dependent, diagonalization procedure used in \cite{qhd3}.
This new diagonalization scheme produces a considerably simpler representation of
QHD-III than the representation given in \cite{qhd3}.  It is shown how, to
${\cal{O}}(g_\rho^2)$, the new diagonalization procedure is equivalent to
simply rescaling the pion field by the ratio $m_\rho / m_a$.

To demonstrate the need for massive vector mesons, consider the
$\sigma-\omega$ model with pions (QHD-II) \cite{qhd3}:
\begin{eqnarray}\label{sigma}
\cal{L}_{\sigma-\omega} &=&\bar{\psi}[i\gamma^{\mu}(\partial_{\mu} + ig_v V_{\mu}) -
g_{\pi}(s+i\gamma_5  \mbox{\boldmath $\tau$} \cdot \mbox{\boldmath $\pi$})]\psi +
\frac{1}{2}(\partial_{\mu}s \partial^{\mu}s + \partial_{\mu}
\mbox{\boldmath $\pi$} \cdot \partial^{\mu} \mbox{\boldmath $\pi$} )\nonumber \\
& & -\frac{1}{4}\lambda(s^2+ \mbox{\boldmath $\pi$}^2 - v^2)^2 -
\frac{1}{4}F_{\mu\nu}F^{\mu \nu} + \frac{1}{2}m_v^2V_{\mu}V^{\mu} +
\epsilon s.
\end{eqnarray}

In equation (\ref{sigma}), $V^\mu$ represents the $\omega$ field, and
$\epsilon s$ is the chiral symmetry violating term that gives a mass
to the pion. 
The global $SU(2)_L\times SU(2)_R$ symmetry of
${\cal{L}}_{\sigma-\omega}$ is now made local, and the scalar field is 
given a vacuum expectation value ($s=\sigma_\circ - \sigma$
with $\sigma_\circ\equiv M/g_{\pi}$).  This yields for the lagrangian,
${\cal{L}}_g$, of the gauged
$\sigma-\omega$ model with pions:
\begin{eqnarray}\label{gauged}
{\cal{L}}_g&=&\bar{\psi}\{i\gamma^{\mu}[\partial_{\mu} +
ig_v V_{\mu} +\frac{i}{2}g_\rho\mbox{\boldmath $\tau$}\cdot(\mbox{\boldmath $\rho$}_\mu + \gamma_5\mbox{\boldmath $a$}_\mu)] -
(M-g_{\pi}\sigma)-ig_\pi\gamma_5  \mbox{\boldmath $\tau$} \cdot \mbox{\boldmath $\pi$}^\prime\}\psi \nonumber\\
& &+\frac{1}{2}[(\partial_{\mu} \mbox{\boldmath $\pi$}^\prime +
g_{\rho}\sigma \mbox{\boldmath $a$}_\mu 
+ g_{\rho} \mbox{\boldmath $\pi$}^\prime \times \mbox{\boldmath $\rho$}_{\mu})^2 - m_{\pi^\prime}^2
\mbox{\boldmath $\pi$}^\prime \cdot \mbox{\boldmath $\pi$}^\prime] + \frac{1}{2}[(\partial_\mu\sigma -
g_{\rho}  \mbox{\boldmath $\pi$}^\prime
\cdot \mbox{\boldmath $a$}_\mu )^2 - m_{\sigma}^2\sigma^2]\nonumber\\ 
& & - g_{\rho} \sigma_\circ
\mbox{\boldmath $a$}^{\mu}\cdot (\partial_\mu\mbox{\boldmath $\pi$}^\prime + g_\rho\sigma \mbox{\boldmath $a$}_\mu 
+ g_\rho  \mbox{\boldmath $\pi$}^\prime \times \mbox{\boldmath $\rho$}_{\mu})\nonumber\\
& &  +
\frac{m_{\sigma}^2 - m_{\pi^\prime}^2}{2\sigma_\circ}
\sigma\left(\sigma^2 +  {\mbox{\boldmath $\pi$}^\prime}^2\right) - 
\frac{m_{\sigma}^2 - m_{\pi^\prime}^2}{8\sigma_\circ^2}\left(\sigma^2 +
 {\mbox{\boldmath $\pi$}^\prime}^2\right)^2- \frac{1}{4}F_{\mu\nu}F^{\mu \nu} +
\frac{1}{2}m_v^2V_{\mu}V^{\mu} \nonumber\\
& &
  - \frac{1}{4}\mbox{\boldmath $R$}_{\mu\nu}\cdot \mbox{\boldmath $R$}^{\mu \nu}
- \frac{1}{4}\mbox{\boldmath $A$}_{\mu\nu}\cdot \mbox{\boldmath $A$}^{\mu \nu} 
+\frac{1}{2} g_\rho^2\sigma_\circ^2 \mbox{\boldmath $a$}_\mu\cdot \mbox{\boldmath $a$}^\mu.
\end{eqnarray}

From equation (\ref{gauged}), we see that the SSB in the $\sigma$ sector has given a mass $M$
to the nucleon and a mass $g_\rho\sigma_\circ$ to the $\mbox{\boldmath $a_1$}$.  The
$\sigma$ mass is $m_\sigma$, and the $\mbox{\boldmath $\rho$}$ remains massless.  We also 
note the presence of a bilinear term, $-g_\rho\sigma_\circ
\mbox{\boldmath $a$}^\mu\cdot\partial_\mu\mbox{\boldmath $\pi$}^\prime$, and thus the need for diagonalization.  Most
importantly, the pion has disappeared and has been replaced by the auxiliary
field, $\mbox{\boldmath $\pi$}^\prime$.  The fact that $\mbox{\boldmath $\pi$}^\prime$ is an auxiliary
field can be seen either by counting the degrees 
of freedom before and after the SSB, or by making the change of
variables:
\begin{equation}\label{aux}
\mbox{\boldmath $a$}_\mu \rightarrow \mbox{\boldmath $a$}_\mu + \frac{1}{g_\rho\sigma_\circ}\partial_\mu 
\mbox{\boldmath $\pi$}^\prime.
\end{equation}

This change of variables both diagonalizes the lagrangian and forces a 
cancelation of the kinetic energy term,
$\partial_\mu\mbox{\boldmath $\pi$}^\prime\cdot\partial^\mu\mbox{\boldmath $\pi$}^\prime$, clearly identifying
$\mbox{\boldmath $\pi$}^\prime$ as an auxiliary field.  The initial pion was ``eaten'' by
the $\mbox{\boldmath $a_1$}$.  ${\cal{L}}_g$ can also be diagonalized using a
gauge-fixing function and working in a $\xi$ gauge \cite{peskin}.  For simplicity,
take
the chiral limit ($m_{\pi^\prime}=0$) and
add to the lagrangian the gauge-fixing function, $-\frac{1}{2}G^2$,  where:
\begin{equation}\label{g}
G=\frac{1}{\sqrt{\xi}}(\partial_\mu \mbox{\boldmath $a$}^\mu + \xi g_\rho\sigma_\circ
\mbox{\boldmath $\pi$}^\prime ).
\end{equation}

The gauge-fixed lagrangian, ${\cal{L}}_g^{gf}$, becomes:
\begin{equation}\label{gf}
{\cal{L}}_g^{gf} = {\cal{L}}_g - g_\rho\sigma_\circ\mbox{\boldmath $\pi$}^\prime
\cdot\partial_\mu \mbox{\boldmath $a$}^\mu -
\frac{1}{2\xi}(\partial_\mu \mbox{\boldmath $a$}^\mu)^2 -\frac{1}{2}\xi g_\rho^2\sigma_\circ^2 
{\mbox{\boldmath $\pi$}^\prime}^2.
\end{equation}

It is noted that the second term in (\ref{gf}) cancels the bilinear term
in (\ref{gauged}) after a partial integration, and that the
$\mbox{\boldmath $\pi$}^\prime$ has acquired a mass that depends on the gauge parameter,
$\xi$.  From (\ref{gf}), the propagators in momentum space of the
$\mbox{\boldmath $a$}_\mu$ and the $\mbox{\boldmath $\pi$}^\prime$ are found to be respectively:
\begin{equation}\label{aprop}
\Delta_{\mu\nu}^{ij} = \frac{-i\delta^{ij}}{q^2 - m_a^2}[g_{\mu\nu} -
\frac{q_\mu q_\nu}{m_a^2}] - \frac{i\delta^{ij}}{q^2 - \xi
m_a^2}\frac{q_\mu q_\nu}{m_a^2},
\end{equation}
\begin{equation}\label{pprop}
\Delta^{ij} = \frac{i\delta^{ij}}{q^2 - \xi m_a^2}.
\end{equation}

Here,  $m_a=g_\rho\sigma_\circ$ and  we have separated
out the $\xi$-dependent part of the $\mbox{\boldmath $a$}_\mu$ propagator in the last
term of (\ref{aprop}).  In the limit
$\xi \rightarrow \infty$, the $\mbox{\boldmath $\pi$}^\prime$ decouples from the problem, while the
$\mbox{\boldmath $a$}_\mu$ propagator goes into the unitary gauge.  In an arbitrary $\xi$ 
gauge, the $\xi$-dependent part of the $\mbox{\boldmath $a$}_\mu$ propagator always
cancels the contribution coming from $\mbox{\boldmath $\pi$}^\prime$ exchange.  This 
can be seen in nucleon-nucleon scattering at tree level in figure (I)
\cite{peskin}.  The $\mbox{\boldmath $\pi$}^\prime$ exchange diagram is precisely canceled
by the $\xi$-dependent part of the $\mbox{\boldmath $a_1$}$ propagator as is easily
verified.  The fact that the $\mbox{\boldmath $\pi$}^\prime$ does not contribute to
physical processes, and the fact that the 
gauge invariance is preserved, are visibly true because the $\mbox{\boldmath $\pi$}^\prime$ has
the correct mass: $m_{\pi^\prime}^2 = \xi m_a^2$.\footnote{Since the
non-linear $\sigma$ model is the limit of the linear $\sigma$ model as
$m_\sigma\rightarrow\infty$, the pion also disappears in the gauged non-linear
$\sigma$ model.}

\begin{center}
\setlength{\unitlength}{1cm}
\begin{picture}(11.0,4.0)(0,0)\label{NN}
\ArrowLine(57,57)(28,85)
\ArrowLine(28,28)(57,57)
\ZigZag(57,57)(113,57){4}{6}
\ArrowLine(113,57)(142,85)
\ArrowLine(142,28)(113,57)
\Text(0.7,3.0)[]{$N$}
\Text(0.7,1.0)[]{$N$}
\Text(5.2,3.0)[]{$N$}
\Text(5.2,1.0)[]{$N$}
\Text(3.0,2.5)[]{$\mbox{\boldmath $a_1$}$}
\Text(5.5,2.0)[]{+}
\ArrowLine(198,57)(170,85)
\ArrowLine(170,28)(198,57)
\DashLine(198,57)(255,57){4}
\ArrowLine(255,57)(283,85)
\ArrowLine(283,28)(255,57)
\Text(5.8,3.0)[]{$N$}
\Text(5.8,1.0)[]{$N$}
\Text(10.2,3.0)[]{$N$}
\Text(10.2,1.0)[]{$N$}
\Text(8.0,2.5)[]{$\mbox{\boldmath $\pi$}^\prime$}
\end{picture}\\
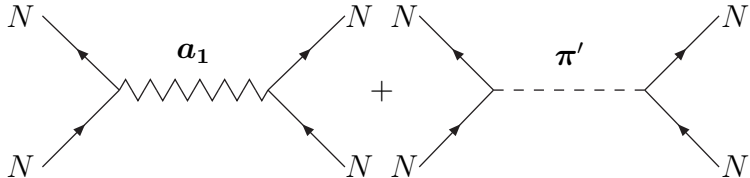
Figure (I): {\sl $\mbox{\boldmath $a_1$}$ and $\mbox{\boldmath $\pi$}^\prime$ exchange diagrams in NN scattering.}
\end{center}

To retain the physical pion in the model, the gauge bosons must
develop a mass from {\it outside} the $\sigma$ sector.  In QHD-III
\cite{qhd3}, a Higgs sector composed of left and right complex doublets
is included to preserve the gauge invariance of the model:

\begin{eqnarray}\label{h}
{\cal{L}}_H &=& \partial_\mu\phi_R^\dagger\partial^\mu\phi_R +
\partial_\mu\phi_L^\dagger\partial^\mu\phi_L + \mu_H^2(\phi_R^\dagger
\phi_R +\phi_L^\dagger \phi_L)\\ \nonumber
& &-\frac{\lambda_H}{4}[(\phi_R^\dagger\phi_R)^2 + (\phi_L^\dagger \phi_L)^2].
\end{eqnarray}

As detailed in \cite{qhd3}$, {\cal{L}}_H$ is then made locally-invariant by
minimal substitution.  After the SSB in the full theory, and the elimination of the
Goldstone bosons from the Higgs sector by going into the unitary
gauge, the Higgs lagrangian becomes:
\begin{eqnarray}\label{higgs}
{\cal{L}}_H &=& \frac{1}{2}(\partial_\mu\eta\partial^\mu\eta -
m_H^2\eta^2) + \frac{1}{2}(\partial_\mu\zeta\partial^\mu\zeta -
m_H^2\zeta^2)\\ \nonumber
& & \frac{1}{2}[g_\rho m_\rho \eta +
\frac{1}{4}g_\rho^2(\eta^2+\zeta^2)] (\mbox{\boldmath $\rho$}_\mu\cdot\mbox{\boldmath $\rho$}^\mu +
\mbox{\boldmath $a$}_\mu\cdot \mbox{\boldmath $a$}^\mu) + (g_\rho m_\rho\zeta +
\frac{1}{2}g_\rho^2\eta\zeta) \mbox{\boldmath $\rho$}_\mu\cdot \mbox{\boldmath $a$}^\mu \\ \nonumber
& & -(\frac{3m_H^2 g_\rho}{4m_\rho}\eta +\frac{3m_H^2
g_\rho^2}{16m_\rho^2}\eta^2)\zeta^2 -\frac{m_H^2
g_\rho}{4m_\rho}\eta^3 -\frac{m_H^2
g_\rho^2}{32m_\rho^2}(\eta^4 + \zeta^4)\\ \nonumber
& &+\frac{1}{2}m_\rho^2\mbox{\boldmath $\rho$}_\mu\cdot \mbox{\boldmath $\rho$}^\mu + \frac{1}{2} m_\rho^2 \mbox{\boldmath $a$}_\mu
\cdot \mbox{\boldmath $a$}^\mu,
\end{eqnarray}

where:
\begin{equation}
\mu_H^2=\frac{1}{2}m_H^2, \hspace{35pt} u^2 =
\frac{8\mu_H^2}{\lambda_H} = \frac{4m_\rho^2}{g_\rho^2}, \hspace{35pt} 
\lambda_H = \frac{m_H^2 g_\rho^2}{m_\rho^2}.
\end{equation}

In the equation above, $u$ is twice the vacuum expectation value of the scalar fields in
the Higgs sector.  The Higgs fields $\eta$ and $\zeta$ are
respectively scalar and pseudoscalar fields.  The QHD-III lagrangian,
${\cal{L}}_{III}$, is given by:
\begin{equation}\label{l3}
{\cal{L}}_{III} = {\cal{L}}_g + {\cal{L}}_H
\end{equation}

It is seen that the $\mbox{\boldmath $a_1$}$ obtains contributions to its mass, $m_a$, from both the 
$\sigma$ sector and from the Higgs sector.  Thus, the $\mbox{\boldmath $a_1$}$ comes out
naturally more massive than the $\mbox{\boldmath $\rho$}$ with:
\begin{equation}\label{ma}
m_a^2 = m_\rho^2 + g_\rho^2\sigma_\circ^2 > m_\rho^2
\end{equation}

The pion in QHD-III is now {\it real} and ${\cal{L}}_{III}$ must be
diagonalized further to remove the term $-g_\rho\sigma_\circ
\mbox{\boldmath $a$}^\mu\cdot\partial_\mu\mbox{\boldmath $\pi$}^\prime$ in ${\cal{L}}_g$.  This is achieved by
performing the change of variables:
\begin{equation}\label{change}
\mbox{\boldmath $a$}_\mu \rightarrow \mbox{\boldmath $a$}_\mu + \frac{g_\rho\sigma_\circ}{m_a^2}
\partial_\mu \mbox{\boldmath $\pi$}^\prime, \hspace{40pt} \mbox{\boldmath $\pi$}^\prime = \frac{m_a}{m_\rho}\mbox{\boldmath $\pi$},
\hspace{40pt} m_{\pi^\prime} = \frac{m_\rho}{m_a}m_\pi.
\end{equation}

$\mbox{\boldmath $\pi$}$ is now the physical pion field.  Since the $\mbox{\boldmath $a_1$}$ appears in many
places in ${\cal{L}}_{III}$, and the
change of variables involves the gradient of the pion, the
diagonalized lagrangian is quite complicated with momentum-dependent
vertices showing up everywhere.

In contrast, the gauge invariance of QHD-III allows another, simpler
diagonalization procedure to work.  Consider equation (\ref{higgs}): to
obtain ${\cal{L}}_H$, the Goldstone
bosons of the Higgs sector were eliminated from the theory by going into 
the unitary gauge, as is done in the standard model: these Goldstone
bosons are fictitious since they can be removed by a gauge
transformation.  This can be seen by counting the degrees of freedom:
initially, before the SSB in the Higgs sector, there are two complex
doublets (the Higgs fields) which yield eight degrees of freedom,
and two massless isovector fields (the $\mbox{\boldmath $\rho$}$ and the $\mbox{\boldmath $a_1$}$)
which add 12 degrees of freedom; this yields a total of 20 degrees of freedom.
After the SSB, there are two isoscalar Higgs fields (the $\eta$ and
the $\zeta$), and two massive
isovector fields for a total 20 degrees of
freedom.  Thus, two isovector fields  ``disappeared'' from the
Higgs sector to become the longitudinal polarization
states of the $\mbox{\boldmath $\rho$}$ and the $\mbox{\boldmath $a_1$}$; these are the Goldstone bosons.  If one does not work in the
unitary gauge, the Goldstone bosons must still couple to the other
fields to maintain gauge invariance.  One of the isovector fields must
couple directly to the $\mbox{\boldmath $\rho$}$ and must therefore be a scalar field
because of parity conservation,
while the other isovector field which we denote as $\mbox{\boldmath $\chi$}^\prime$ must
couple directly to the $\mbox{\boldmath $a_1$}$ and must be a pseudoscalar field.  The scalar isovector
field that was eaten by the $\mbox{\boldmath $\rho$}$ is not needed for
this discussion, and can be decoupled from the problem independently of the other
Goldstone bosons.  As for the pseudoscalar Goldstone bosons, the
contributions to ${\cal{L}}_H$ stemming from
working in an arbitrary gauge, and therefore keeping the
$\mbox{\boldmath $\chi$}^\prime$ are:
\begin{equation}\label{chi}
{\cal{L}}_H^\prime = {\cal{L}}_H + \partial_\mu
\mbox{\boldmath $\chi$}^\prime \cdot \partial^\mu \mbox{\boldmath $\chi$}^\prime + m_\rho \mbox{\boldmath $a$}_\mu \cdot
\partial^\mu \mbox{\boldmath $\chi$}^\prime + \Delta{\cal{L}}_H^\prime.
\end{equation}

$\Delta{\cal{L}}_H^\prime$ represents terms given in the appendix for
completeness.  The presence of
the bilinear term in equation (\ref{chi}) is noted and is typically
removed with a gauge-fixing function similar to the one given in
(\ref{g}).  The $\mbox{\boldmath $\chi$}^\prime$ is then decoupled by taking $\xi$ to
infinity as discussed in the case of the $\mbox{\boldmath $\pi$}^\prime$ below equation
(\ref{pprop}).  This is what was done in \cite{qhd3}.

Consider instead the following gauge-fixing function:
\begin{equation}\label{ga}
G_a=\frac{1}{\sqrt{\xi}}(\partial_\mu \mbox{\boldmath $a$}^\mu - \xi m_\rho \mbox{\boldmath $\chi$}^\prime +
\xi g_\rho\sigma_\circ \mbox{\boldmath $\pi$}^\prime )
\end{equation}

Adding $-\frac{1}{2}G_a^2$ to ${\cal{L}}_{III} = {\cal{L}}_H^\prime +
{\cal{L}}_g$ cancels the bilinear terms: $-g_\rho\sigma_\circ
\mbox{\boldmath $a$}^\mu\cdot\partial_\mu\mbox{\boldmath $\pi$}^\prime$ and $m_\rho \mbox{\boldmath $a$}_\mu \cdot
\partial^\mu \mbox{\boldmath $\chi$}^\prime$.  The propagator of the $\mbox{\boldmath $a_1$}$ is exactly as 
in equation (\ref{aprop}) with $m_a$ now given by (\ref{ma}).  What is
left from the gauge-fixing function is:
\begin{equation}\label{chipi}
-\frac{1}{2}G_a^2 \doteq -\frac{1}{2}\xi m_\rho^2 \mbox{\boldmath $\chi$}^\prime \cdot \mbox{\boldmath $\chi$}^\prime
-\frac{1}{2} \xi g_\rho^2 \sigma_\circ^2 \mbox{\boldmath $\pi$}^\prime \cdot \mbox{\boldmath $\pi$}^\prime +
\xi g_\rho\sigma_\circ m_\rho \mbox{\boldmath $\pi$}^\prime \cdot \mbox{\boldmath $\chi$}^\prime.
\end{equation}

Equation (\ref{chipi}) needs a further diagonalization to cancel the
last term; this can be achieved by making the following change of
variables:
\begin{equation}\label{finald}
\mbox{\boldmath $\chi$}^\prime = a\mbox{\boldmath $\pi$} + b\mbox{\boldmath $\chi$}; \hspace{45pt} \mbox{\boldmath $\pi$}^\prime = c\mbox{\boldmath $\pi$} + d\mbox{\boldmath $\chi$}.
\end{equation}

The parameters \{$a,b,c,d$\} are constrained by the requirements that
all bilinear terms that couple the $\mbox{\boldmath $\pi$}$ and the $\mbox{\boldmath $\chi$}$ be canceled, and
that the kinetic energies be normalized:
\begin{equation}\label{norm}
\partial_\mu \mbox{\boldmath $\pi$}^\prime \cdot \partial^\mu \mbox{\boldmath $\pi$}^\prime + \partial_\mu
\mbox{\boldmath $\chi$}^\prime \cdot \partial^\mu \mbox{\boldmath $\chi$}^\prime = \partial_\mu
\mbox{\boldmath $\pi$} \cdot \partial^\mu \mbox{\boldmath $\pi$} + \partial_\mu
\mbox{\boldmath $\chi$} \cdot \partial^\mu \mbox{\boldmath $\chi$}
\end{equation}

These equations have the solution:
\begin{equation}\label{abcd}
a=\frac{g_\rho\sigma_\circ}{m_a}, \hspace{25pt} b= \frac{m_\rho}{m_a}, 
\hspace{25pt} c=\frac{m_\rho}{m_a},\hspace{25pt} d=-
\frac{g_\rho\sigma_\circ}{m_a}
\end{equation}

It is found that the masses of the $\mbox{\boldmath $\chi$},\mbox{\boldmath $\pi$}$ fields are respectively:
\begin{equation}\label{pimass}
m_\chi^2 = \xi (m_\rho^2 + g_\rho^2 \sigma_\circ^2) = \xi m_a^2,
\hspace{45pt} m_\pi = \frac{m_\rho}{m_a} m_{\pi^\prime}.
\end{equation}

The mass of the $\mbox{\boldmath $\chi$}$ is exactly the mass needed to cancel the
$\xi$-dependent part of the $\mbox{\boldmath $a_1$}$ propagator given in equation
(\ref{aprop}), as discussed below equation (\ref{pprop}).  The $\mbox{\boldmath $\chi$}$
is the field that provided the longitudinal
polarization states of the $\mbox{\boldmath $a_1$}$ after the SSB.

$\mbox{\boldmath $\chi$}$ can be decoupled by taking $\xi \rightarrow \infty$ and
equation (\ref{finald}) becomes:
\begin{equation}\label{scale}
\mbox{\boldmath $\chi$}^\prime = \frac{g_\rho\sigma_\circ}{m_a} \mbox{\boldmath $\pi$}; \hspace{45pt}
\mbox{\boldmath $\pi$}^\prime = \frac{m_\rho}{m_a}\mbox{\boldmath $\pi$}.
\end{equation}

It is thus seen that the diagonalization is achieved by a {\it constant
rescaling} of the pion field.  Because of the constraint (\ref{norm}), 
the pion kinetic energy is not rescaled.  This completely avoids the
introduction of new momentum-dependent vertices.

Through the first equation 
of (\ref{scale}), the pion couples directly to the Higgs
fields.\footnote{This was also the case in the change of variable
(\ref{change}) as is
seen by substituting (\ref{change}) into equation (\ref{higgs}).}  The
pion-Higgs vertices generally involve a high power of $g_\rho$ as seen
in the appendix.  By inspection, it is seen that any amplitude that does
not involve both a Higgs field and a gauge boson as external legs will
not contribute to ${\cal{O}}(g_\rho^2)$.  Hence, to order
${\cal{O}}(g_\rho^2)$, all S-matrix elements that do not involve a
Higgs field as either an incoming or an outgoing field, can be
calculated by ignoring the Higgs sector, and rescaling the pion field
by the constant factor: $m_\rho/m_a$.   In the chiral limit, when
$m_\pi=0$, this new, simpler representation of the QHD-III
lagrangian is completely equivalent to the one given in \cite{qhd3};
when $m_\pi=0$,
the two representations lead to exactly the same physical predictions.  In 
particular, a calculation in QHD-III of an amplitude such as $\mbox{\boldmath $\pi$}\mbox{\boldmath $\pi$}$
scattering to one loop in the chiral limit
\cite{paper} is considerably simplified.

When $m_\pi \not= 0$, the two representations will  differ slightly in
their physical predictions since the pion mass term violates chiral
symmetry; i.e. the contribution of the chiral symmetry breaking term is
gauge dependent.  Although the pion mass
term is gauge-dependent, the symmetry breaking term in equation (\ref{sigma}) is always chosen
so as to result in the physical pion mass once the gauge has been
picked.  This is what was done in equation (\ref{pimass}).

In {\it summary}, it is shown that a quantum field model of the strong
interactions, based on an $SU(2)_L \times SU(2)_R$ locally-invariant
$\sigma$ model, is not realistic if the corresponding gauge bosons are
massless before the SSB; in such a model, the pion disappears.  We saw
how putting masses by hand violates gauge invariance and forces a
complicated diagonalization procedure on the model. 
QHD-III is reviewed, and the gauge invariance of the model is exploited to
provide a {\it new}, considerably {\it simpler} representation of the model that
makes it more accessible.  The pion here appears as a physical degree
of freedom.

The author would like to thank J.D. Walecka for his guidance and also
C. Carlson for useful discussions.  This work was supported under
U.S. DOE Grant No. DE-FG0297ER41023.

\section{appendix}

When retaining the Goldstone boson $\mbox{\boldmath $\chi$}^\prime$ in the model, the
following extra terms appear:
\begin{eqnarray}\label{chihiggs}
\Delta{\cal{L}}_H^\prime &=& \frac{g_\rho}{2} ( \eta \mbox{\boldmath $a$}_\mu \cdot
\partial^\mu \mbox{\boldmath $\chi$}^\prime + \zeta \mbox{\boldmath $\rho$}_\mu \cdot \mbox{\boldmath $\chi$}^\prime -
\mbox{\boldmath $\chi$}^\prime \cdot \mbox{\boldmath $a$}_\mu \partial^\mu \eta - \mbox{\boldmath $\chi$}^\prime \cdot
\mbox{\boldmath $\rho$}_\mu \partial^\mu \zeta ) + \frac{g_\rho}{2}\mbox{\boldmath $\chi$}^\prime \times \mbox{\boldmath $\rho$}_\mu \cdot \partial^\mu
\mbox{\boldmath $\chi$}^\prime \\ \nonumber
& &- \frac{g_\rho m_H^2}{4m_\rho} \eta {\mbox{\boldmath $\chi$}^\prime}^2 +
\frac{g_\rho^2}{8} {\mbox{\boldmath $\chi$}^\prime}^2 ( \mbox{\boldmath $a$}_\mu^2 +
\mbox{\boldmath $\rho$}_\mu^2)-\frac{g_\rho^2 m_H^2}{16 m_\rho^2}{\mbox{\boldmath $\chi$}^\prime}^2 (\zeta^2
+ \eta^2)-\frac{g_\rho^2 m_H^2}{32 m_\rho^2}{\mbox{\boldmath $\chi$}^\prime}^4.
\end{eqnarray}

When $\frac{g_\rho\sigma_\circ}{m_a}\mbox{\boldmath $\pi$}$ is substituted for the $\mbox{\boldmath $\chi$}^\prime$ in equation
(\ref{chihiggs}), it is seen that the interactions in
(\ref{chihiggs}) are at least of order $g_\rho^2$.  In particular, 
the terms in the parentheses of the first line of (\ref{chihiggs}) are the only
ones of ${\cal{O}}(g_\rho^2)$, and they all involve the Higgs
fields; the contributions of these terms can be minimized by making
the Higgs fields very heavy.

\end{document}